\documentclass[aps,prb,showpacs,twocolumn,amsmath,amssymb]{revtex4}

\usepackage{amsmath}    
\usepackage{graphicx}   
\usepackage{graphics}
\usepackage{dcolumn}
\usepackage{bm}

\usepackage{multirow}

\usepackage{latexsym}
\usepackage{amssymb}
\usepackage{amsmath}

\usepackage{color}
\usepackage{epstopdf}

\usepackage[normalem]{ulem}
\newcommand{\vd}[1]{\textcolor{red}{#1}}

\begin{document}
\title{An infinite family of $bc8$-like metastable phases in silicon}
\author{Vladimir E. Dmitrienko\footnote{email: dmitrien@crys.ras.ru} and Viacheslav A. Chizhikov\footnote{email: chizhikov@crys.ras.ru}}
\affiliation{A.V. Shubnikov Institute of Crystallography, FSRC ``Crystallography
and Photonics'' RAS, Leninskiy Prospekt 59, 119333, Moscow, Russia}


\begin{abstract} 
We show that new silicon crystalline phases, observed in the experiment
with the laser-induced microexplosions in silicon crystals (Rapp {\it
et al.} Nat. Commun. \textbf{6}, 7555 (2015)), are all superstructures of
a disordered high-symmetry phase with $Ia\bar{3}d$ cubic space group, as
well as known for many years phases $bc8$ (Si-III) and $r8$ (Si-XII). The
physics of this phenomenon is rather nontrivial: The $bc8$-like
superstructures appear as regularly ordered patterns of switchable atomic
strings, preserving everywhere the energetically favorable tetrahedral
coordination of silicon atoms. The variety of superstructures arises
because each string can be switched between two states independently of
the others. An infinite family of different phases can be obtained this
way and a number of them are considered here in detail. In addition to the
known $bc8$, $bt8$, and $r8$ crystals, 128 tetrahedral structures with 16
(6 phases), 24 (22 phases), and 32 (100 phases) atoms per primitive cell
are generated and studied, most of them are new ones. For the coarse-grain
description of the structures with two possible states of switchable
strings, the black/white (switched/nonswitched) Shubnikov symmetry groups
has been used. An {\it ab initio} relaxation of the atomic positions and
lattice parameters shows that all the considered phases are metastable and
have higher density and energy relative to the $bc8$ phase at the ambient
pressure. A possible scenario for appearance of those phases from the
high-temperature amorphous phase is discussed.

\end{abstract}
\maketitle


\section{Introduction}
\label{sec:intro} 

Despite the fact that our contemporary civilization is now based on one
type of silicon crystals with the diamond structure, the diversity of
crystal forms of this chemical element is not much inferior to that of
carbon. There are numerous exotic structures with tetrahedral
\cite{Crain1985,Fan2015} and more complicated atomic arrangements,
including a big family of high-pressure silicon phases with higher
coordinations \cite{Ackland2001,Mujica2003,Haberl2016}. The silicon phases
are interesting first of all owing to their electronic properties and the
latter are determined by details of the atomic structures and impurities.
Another promising application is related to micro-electromechanical
systems (MEMS) \cite{Markku2015} where the elastic properties of silicon
are used. For all these purposes, one needs to look for new silicon-based
crystals with the hope to find new unusual properties and applications.

One of the exotic tetrahedral structures, the $bc8$ phase,  was discovered
in 1963 \cite{Wentorf1963,Kasper1964}; its name means body-centered cubic
with 8 atoms per  primitive cell (that is 16 atoms per the body-centered
cubic ($bcc$) unit cell, the lattice parameter $a=6.64$ \AA, and the space
group is $Ia\bar{3}$). It is also called Si-III because it  has been found
after Si-I (diamond-like) and Si-II ($\beta$-Sn-like) silicon phases
\cite{Ackland2001,Mujica2003}. All the atoms in $bc8$ are
crystallographically equivalent, they are located at the three-fold axes,
$16c$  positions $(x,x,x)$ of  group $Ia\bar{3}$ ($x\approx 0.1$), and
have non-ideal tetrahedral coordination with one interatomic bond directed
along a three-fold cubic axis ($A$-bond) and three longer bonds
($B$-bonds) in non-symmetric directions. Structurally, the $bc8$ phase can be
considered as the cubic arrangement of atomic strings directed along four
$\langle 111 \rangle$ axes (this point is discussed in detail below). This
phase was first obtained from the high-pressure $\beta$-Sn-like phase
after pressure release \cite{Wentorf1963,Kasper1964} and was found to be
metastable at the ambient conditions. Its structural, thermodynamical, and
electronic properties have been studied in detail for many years
\cite{Joannopoulos1973,Joannopoulos1973b,Yin1984,Bismas1987,Brazhkin1992,Crain1994,Clark1994,Zhang2017}.
In the $bc8$ phase, like in diamond, there are only even-membered rings of
interatomic bonds and the  shortest rings are six-membered.

\begin{figure}[h]
\begin{center}
\includegraphics[width=7cm]{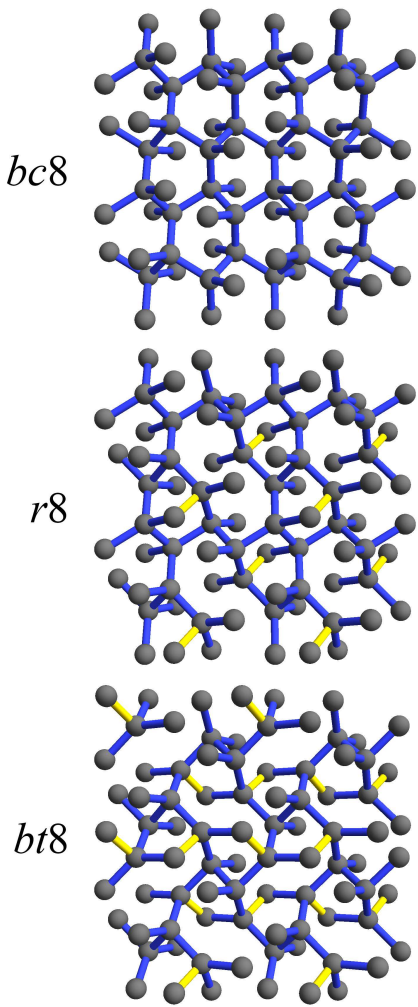}
\caption{\label{fig:bc8-r8-bt8} (Color online) Comparison of the
crystalline silicon structures $bc8$, $r8$, and $bt8$. An atomic layer is
chosen to be parallel to  plane $(100)$ of $bc8$. In the phases $r8$ and
$bt8$, the bonds coinciding with those of $bc8$ are shown in blue, and the
new switched bonds along the $\langle 111\rangle$ directions are yellow.
The most dissimilar phases $bc8$ and $bt8$ differ by switching one
eighth of the total number of bonds.}
\end{center}
\end{figure}

The first $bc8$-like structure, $r8$ or Si-XII,  was discovered in 1994
\cite{Crain1994b}; $r8$ means rhombohedral with 8 atoms per  unit cell,
its space group is $R\bar{3}$, a subgroup of $Ia\bar{3}$. The structure of
$r8$ appears from $bc8$ during a first-order pressure-induced structural
transition as a result of breaking and rebonding of all the $A$-bonds
directed along  cubic direction $[111]$. Indeed, according to
\cite{Crain1994b}, the rebonding can be understood as motion of the atom
at $(x, x, x)$ along  direction $[111]$ of the $bc8$ unit cell until the
$A$-bond to the atom at $(\bar{x}, \bar{x}, \bar{x})$ is broken and a new
$A$-bond to the atom at $(\frac12-x, \frac12-x, \frac12-x)$ is formed
(Fig.~\ref{fig:bc8-r8-bt8}). As a result of the rebonding, all the $[111]$
atomic strings are switched into another sequence of bonds preserving
nevertheless the energetically favorable tetrahedral atomic arrangement.
The phase transition is of the first order because the rebonding is
accompanied by small but finite atomic displacements changing the topology
of rings; in particular, five-membered rings appear in the $r8$ phase.

Both phases $bc8$ and $r8$ can also appear as a result of mechanical
microindentation and  their thermal relaxation back to the diamond silicon
has been studied in numerous research works
\cite{Demishev1996,Domnich2002,Haberl2015,Wong2019}. The electronic and
optical properties of $bc8$ and $r8$  may have interesting practical
applications (see \cite{Malone2008,Wong2019b} and references therein).

In 1999, it was shown \cite{Dmitrienko1999,Dmitrienko1999b} that the
rebonding mechanism found by Crain {\it et al.} \cite{Crain1994b} is
rather universal: in the $bc8$ structure, any set of the atomic strings
can be independently switched without violation of tetrahedral
coordination. In particular, a new tetragonal silicon phase ($bt8$) was
predicted, where all the strings of $bc8$ parallel to two cubic diagonals
(say $[111]$ and $[\bar{1}\bar{1}1]$) were switched and those parallel to
$[1\bar{1}\bar{1}]$ and $[\bar{1}1\bar{1}]$ were not
(Fig.~\ref{fig:bc8-r8-bt8}). The space symmetry of $bt8$ is $I4_1/a$ and
it is not a subgroup of $Ia\bar{3}$: there are new symmetry elements,
fourfold screw axes, relating switched and non-switched strings. The $bt8$
structure demonstrates the maximum density of five-membered bond rings
\cite{Dmitrienko1999b} and its electronic properties are expected to be
rather different from those of $bc8$, $r8$, and diamond phases. Energetics
and structural relaxation of $bt8$ have been studied {\it ab initio} both
for silicon \cite{Dmitrienko1999b} and carbon \cite{Dmitrienko2001}. It
has been shown that at the ambient pressure $bt8$ is more dense than $bc8$
and $r8$ and it becomes energetically more favorable than $bc8$ and $r8$
at pressures above 13 GPa where the $\beta$-tin silicon structure is, in
fact, more favorable than all the $bc8$-like phases.  Later on, in 2013,
the $bt8$ phase was independently reinvented and studied {\it ab
initio} for silicon and germanium \cite{Wang2013} and for carbon as
well \cite{Ishikawa2014} (in the latter case the symmetry was claimed
to be $I4_1$ whereas the calculated atomic coordinates corresponded in
fact to the more symmetric $I4_1/a$ space group). The $bc8$-like
structures were also studied by  algebraic geometry \cite{Kraposhin2008}
and by the high-dimensional projection methods used for quasicrystals and
their approximants \cite{Dmitrienko2000}.

The real breakthrough  happened in 2015 when Rapp {\it et al.}
\cite{Rapp2015}  found evidence for several metastable silicon phases
after ultrashort laser-induced confined microexplosions \cite{Rapp2013} at
the interface between a transparent amorphous silicon dioxide layer
(SiO$_2$) and an opaque single-crystal Si substrate. They  determined
the lattice parameters and possible atomic structures of the following
phases: $bt8$, $st12$ (analog of $st12$ in germanium), two tetragonal
phases with 32 atoms per unit cells, and some others. For  description
of the structures they used an {\it ab initio} random structure search
\cite{Mujica2015}. It should be noted that one of the 32-atoms tetragonal
phases was found independently \cite{Zhu2015} using the ideas of
metadynamics and evolutionary algorithms. The exotic silicon phases like
$bc8$, $r8$, and those new discovered by Rapp {\it et al.} \cite{Rapp2015}
have provided a novel insight into the local structure and properties of
the amorphous phase of silicon
\cite{Joannopoulos1973,Joannopoulos1973b,Dmitrienko1999,Ruffell2009}.

In the present paper we suggest a unified description of all new silicon
crystalline phases (except $st12$) observed by Rapp {\it et al.}
\cite{Rapp2015} and numerous similar phases. Those complicated phases are
shown to be the $bc8$-like structures with periodically switched strings,
like in the simple case of $r8$. As a result, the lattice vectors of those
phases are some periods of $bc8$. In addition to known $bc8$, $r8$, and
$bt8$, we generate a complete set of $bc8$-like phases with 16, 24, and 32
atoms per primitive cells and relax their structures {\it ab initio}.

\section{BC8 structure and string switching}
\label{sec:srs}

\begin{figure}[h]
\begin{center}
\includegraphics[width=7cm]{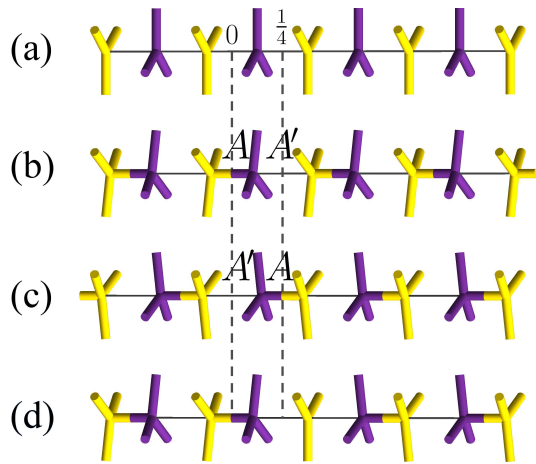}
\caption{\label{fig:string} (Color online)  An atomic string in the structure of $bc8$-like silicon. Each atom has one $A$-bond along the string and three $B$-bonds with  atoms of other strings. The string can be switched from the mean position (a), corresponding to the structure with space group $Ia\bar{3}d$, to one of two extreme states, white (b) and black (c).  Switching from white to black permutes the $A$-bonds  with  $A^\prime$-distances,  preserving tetrahedral atomic arrangement. In addition, the string can be split into white and black parts, separated by defects (d). The switching process may consist in moving such a defect along the string \cite{bondon.gif}.}
\end{center}
\end{figure}

As  mentioned in the introduction, the atomic structure of the $bc8$ phase
can be considered as a set of atomic strings  parallel to the threefold
axes of the cubic space group $Ia\bar{3}$, no. 206. Here the string
structure is described in detail (Fig.~\ref{fig:string}). All  atoms are
located  in the position of $16c$  $(x,x,x)$, $x=x_{w}\approx 0.1$, with
threefold point symmetry, and  each string possesses two nonequivalent
inversion centers  in the positions of $8a$  $(0,0,0)$ and $8b$
$(\frac{1}{4},\frac{1}{4},\frac{1}{4})$. In $bc8$ all strings are
equivalent to each other.  Let us consider one of them, say, that parallel
to axis $[111]$ and passing through the origin $(0,0,0)$.  Then its atoms
have coordinates
$(\frac{n}{2},\frac{n}{2},\frac{n}{2})\pm(x_{w},x_{w},x_{w})$, where $n$
is an arbitrary integer. The $A$-bonds between neighboring atoms of the
string are $(2x_{w},2x_{w},2x_{w})$ and they alternate with  next-neighbor
distances $A^\prime$
$(\frac{1}{2}-2x_{w},\frac{1}{2}-2x_{w},\frac{1}{2}-2x_{w})$. Therefore
the string of atoms looks like a sequence of alternating $A$-bonds
centered at inversion centers
$(\frac{n}{2},\frac{n}{2},\frac{n}{2})$, and approximately 1.5
times longer stretches $A^\prime$ centered at inversion centers
$(\frac{1}{4}+\frac{n}{2},\frac{1}{4}+\frac{n}{2},\frac{1}{4}+\frac{n}{2})$,
(Fig.~\ref{fig:string}b). The string of this type will be called {\it
white}, hence the subindex $w$.

However, there is another value of $x$, $x=x_{b}=\frac{1}{4}-x_{w}$, which
gives an identical $bc8$ structure rotated  relative to the  former by
$\frac{\pi}{2}$ around a twofold axis (the subindex $b$ means {\it
black}). The black and white $bc8$ structures can be also transformed one
to another by small local shifts of atoms along threefold directions
resulting in breaking/switching of $A$-bonds, as suggested
by Crain {\it et al.} \cite{Crain1994b} for the $r8$ phase and described
above in the introduction. In the string picture, the rebonding means
simply that $A^\prime$ and $A$-bonds are locally permuted (switched),
Fig.~\ref{fig:string}c. Thus, for the black $bc8$, in the considered above
string $[111]$ the $A$-bonds are centered at
$(\frac{1}{4}+\frac{n}{2},\frac{1}{4}+\frac{n}{2},\frac{1}{4}+\frac{n}{2})$,
whereas $A^\prime$ at $(\frac{n}{2},\frac{n}{2},\frac{n}{2})$.

An important observation is that any set of  strings can be independently
switched  between black  and white states without violation of
energetically favorable tetrahedral coordination
\cite{Dmitrienko1999,Dmitrienko1999b}. The tetrahedral coordination means
that each atom has four  neighbors, the lengths of bonds are not very
different, and all  interbond angles exceed $\frac{\pi}{2}$. An example of
bond length and angle statistics can be found below in
Section~\ref{sec:microscopic}. The switching of a single isolated string
in the perfect $bc8$  structure has been simulated {\it ab initio}
\cite{Dmitrienko1999b} and it has been found that it costs less then 0.02
eV per atom.

Since every string is allowed  to be switched independently of the others,
there exist  $2^N$ different  combinations of switching, where the total
number of strings $N$ is proportional to the surface area of  a crystal.
This huge number of  possible structures arises from  the purely
combinatorial consideration, without taking into account their  physical
properties. Note that all members of the infinite family of $bc8$-like
crystals have a  similar topological structure. Indeed, the string
switching affects only  the bonds lying along the four axes of type
$\langle 111\rangle$. Any $bc8$-like phase can be obtained from $bc8$ by
switching at most one half of $A$-bonds, i.e. one eighth of the
total number of bonds (Fig.~\ref{fig:bc8-r8-bt8}).

As mentioned above, the rotation of $bc8$ by $\frac\pi 2$ changes the
color of all atomic strings. This means that different  combinations of
switching can define the same structure. For example, the two cases when
all atomic strings  have the same color, ($w,w,w,w$) or ($b,b,b,b$),
correspond to a single structure, namely $bc8$. Further, if the strings
parallel to any two  threefold axes are white, and those parallel to
remaining two axes are black (the cases ($w,w,b,b$), ($w,b,w,b$), {\it
etc}.), then the crystal symmetry becomes tetragonal, and identical $bt8$ 
phases with different orientations of the tetragonal axis are
obtained. Finally, if the sign of the strings parallel to any  threefold
axis is opposite to the sign of other strings (the cases ($w,w,w,b$),
($w,b,b,b$), {\it etc}.) then identical $r8$ phases arise with
different orientations of the rhombohedral axis. The three phases, $bc8$,
$bt8$, and $r8$, exhaust the list of $bc8$-like structures with the
minimal primitive cell containing 8 atoms. Their common feature is that
parallel strings have the same color.

In spite of the obvious connection of the phases $bc8$ and $bt8$, there is no
group-subgroup relation between them.  We can assume the existence of a
supersymmetric phase with space group $Ia\bar{3}d$, which is a supergroup
for both $Ia\bar{3}$ and $I4_1/a$.  An evident way to construct the
superphase  is to turn all the strings into the half-switched state. In
this case, all  atoms  are located in the position of $16b$ of group
$Ia\bar{3}d$ with three-coordinated graphene-like  environment
(Fig.~\ref{fig:string}a).  However,  this environment is not favorable for
the silicon atoms. Another, more physical way to obtain  a supersymmetric
phase is to  switch every string randomly  to one of  two possible states,
white or black  (Figs.~\ref{fig:string}b,~\ref{fig:string}c). In this
case, the $Ia\bar{3}d$ symmetry is a result of disorder
\cite{Dmitrienko1999}, and the phase transition to $bc8$, $r8$, $bt8$,
or a more complicated structure  is of disorder-order type.

\section{BC8-like superstructures with enlarged primitive cells}
\label{sec:enlarged}

A more complicated $bc8$-like phase with lager primitive cell can occur
if the structure contains parallel atomic strings  of different colors.
The primitive cell of  the phase  is a multiple in volume and number of
atoms to the {\it primitive} cell of $bc8$, and  its Bravais lattice is a
subset of the $bcc$ Bravais lattice of $bc8$. In addition, each
Bravais lattice corresponds to several different phases  due to multiple
ways to color the strings passing through its primitive cell. In order to
enumerate and classify the structures with the same Bravais lattice,  we
need to calculate the number of independent, i.e. not connected by
periodicity, strings in each of the four directions $\langle 111 \rangle$.
This number is proportional to the volume of the primitive cell and
inversely proportional to the smallest lattice period along the direction.
For example, the number $N_{111}$ of independent strings parallel to  axis
$[111]$  is equal to the greatest common divisor of three triple products
\begin{equation}
\begin{array}{l}
n_a = (1,1,1) \cdot [\mathbf{b} \times \mathbf{c}] , \\
n_b = (1,1,1) \cdot [\mathbf{c} \times \mathbf{a}] , \\
n_c = (1,1,1) \cdot [\mathbf{a} \times \mathbf{b}] ,
\end{array}
\end{equation}
where $\mathbf{a}$, $\mathbf{b}$, and $\mathbf{c}$ are the  Bravais lattice periods, expressed in the parameters of the initial $bcc$ lattice.

The numbers $N_{\bar{1}\bar{1}1}$, $N_{\bar{1}1\bar{1}}$, and $N_{1\bar{1}\bar{1}}$ can be calculated in the same way. Further, the combinatorial  number of possible  two-colorings of strings is equal to $2^{N_{111} + N_{\bar{1}\bar{1}1} + N_{\bar{1}1\bar{1}} + N_{1\bar{1}\bar{1}}}$, however, the actual number of different phases is significantly less. For example,  in the case of the smallest primitive cell ($N_{111} = N_{\bar{1}\bar{1}1} = N_{\bar{1}1\bar{1}} = N_{1\bar{1}\bar{1}} = 1$),  sixteen possible combinations of  switching  define only  three different phases, $bc8$, $bt8$, and $r8$. This significant reduction in the number of different structures is due to two reasons. First, some  structures can be connected by the elements of  group $Ia\bar{3}d$  not included in the  own space group of the structures. Second, some combinations of switching can define structures with smaller primitive cells.

\begin{figure}[h]
\begin{center}
\includegraphics[width=7cm]{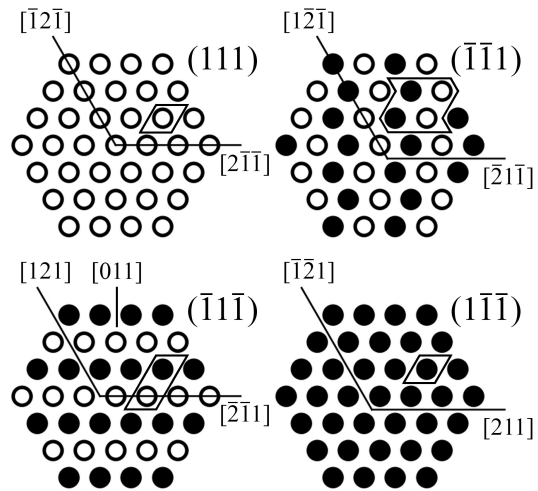}
\caption{\label{fig:m32-strings} Atomic strings in  phase $m32$ ($m32$-8)  in projections onto the planes $\{111\}$ perpendicular to them. For each projection its 2D generating cell is shown. White and black circles indicate the switching of strings. For projections $(\bar{1}1\bar{1})$ and $(\bar{1}\bar{1}1)$ the string switching alternates along the perpendicular directions $[011]$ and $[\bar{2}1\bar{1}]$, correspondingly.}
\end{center}
\end{figure}

\begin{figure}[h]
\begin{center}
\includegraphics[width=7cm]{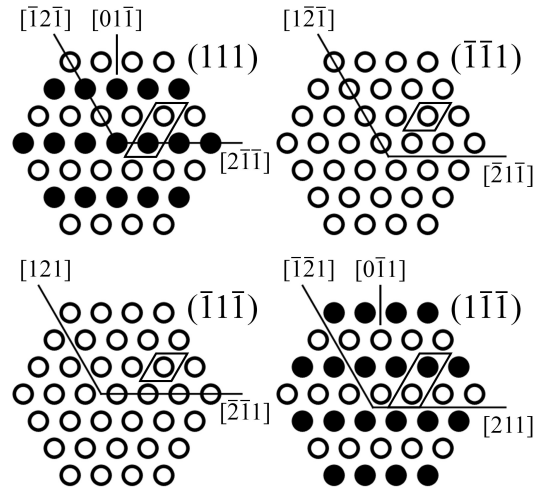}
\caption{\label{fig:m16-strings} Atomic strings in  phase $m32^\star$ ($sm16$-2).  For projections $(111)$ and $(1\bar{1}\bar{1})$ the string switching alternates along the same direction $[01\bar{1}]$.}
\end{center}
\end{figure}

\begin{figure}[h]
\begin{center}
\includegraphics[width=7cm]{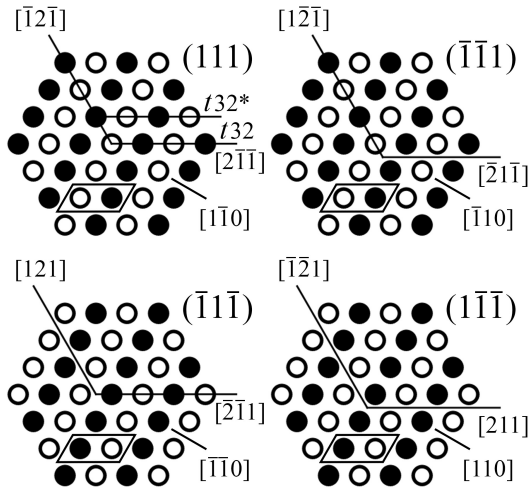}
\caption{\label{fig:t32-strings} Atomic strings in  phases $t32$ ($t32$-1) and $t32^\star$ ($t32$-3). Both phases are characterized by alternation of the string switching along the perpendicular directions $[110]$ and $[1\bar{1}0]$.  Phase $t32^\star$ differs from $t32$ by the inverse switching of all the strings  parallel to axis $[111]$.}
\end{center}
\end{figure}

Table~\ref{tab:phases} shows the currently known $bc8$-like phases of silicon. Recently, several new  structures were experimentally observed, with primitive cells two and four times larger than that of $bc8$ \cite{Rapp2015}. They are listed in the bottom part of  the table, with the names  being written according to Ref.~\cite{Rapp2015}  as well as in our  own notation (in parentheses).  Along with conventional space groups of the crystals, we also indicate the black and white (Shubnikov) groups,  the meaning of which will be explaned later in Section~\ref{sec:magnetic}. The last column shows  the unit cell periods  expressed in the parameters of the initial $bcc$ lattice.  The phases listed in Table~\ref{tab:phases}, unlike some others (diamond, $st12$, etc.), belong to the same family and can be described in the language of string switching, Figs.~\ref{fig:m32-strings}--\ref{fig:t32-strings}.

The list from Table~\ref{tab:phases} is far from being exhaustive. As a
first goal, we would like to find all the similar structures with double,
triple, and quadruple cells,  containing 16, 24, and 32 silicon atoms,
respectively.  First we find all the  Bravais lattices with  the specified
cell volumes,  which are subsets of the initial $bcc$ lattice. Then,  all
possible structures  are obtained by the enumeration of two-colorings of
the independent strings.

In order to avoid double counting, we discard the structures  with
smaller primitive cells.  For example, all the structures with  basis
vectors $(100)$, $(010)$, and $(001)$ (lattice 16-2 in
Table~\ref{tab:cells}) actually have a two times smaller primitive cell,
and therefore they coincide with the phases $bc8$, $bt8$, and $r8$. Thus, it
turns out that some lattices  generate no new phases.  It is also possible
that different combinations of  switching  define the structures connected
to each other by symmetry transformations. Such identical structures  are
also discarded. The statistical results of the search are summarized in
Table~\ref{tab:cells}. Seven lattices (one with double, two with triple,
and four with quadruple  primitive cells) give rise to 128 structures (or
177 if we consider  chiral enantiomorphs as different  phases), four of
which are listed in Table~\ref{tab:phases}. The supplemental material
contains the  list of all the new phases  and  crystallographic
information files (CIFs) with atomic coordinates \cite{description}.

\section{The magnetic analogy}
\label{sec:magnetic}

 Let us define the operation of {\it conjugation} as the simultaneous switching of all strings.  As mentioned above, for the phases $bc8$, $bt8$, and $r8$ the conjugation is equivalent to a rotation of the crystal as a whole and it does not lead to  appearance of a new phase. The question arises whether this is a common property of   the $bc8$-like structures? The answer is no, and already among the crystals with a double primitive cell we find two different triclinic structures, $a16$-3 and $a16$-4, conjugated to each another. Note that this operation does not affect the elements of spatial symmetry and therefore both phases  are of the same space group. Nevertheless, their physical properties (energy, atomic density, cell parameters, etc.), generally speaking, should differ.

In order to prove that conjugated structures have the same space group, an analogy with magnetic crystals can be suggested. Let us  assign to each  atom  the shift vector from  a symmetric graphene-like position  of  group $Ia\bar{3}d$ to its real position with tetrahedral coordination. This vector is always parallel to the threefold axis (string) passing through the atom, whereas its direction alternates along the  string. The situation resembles the ordering of magnetic moments arranged in the nodes of the original  $Ia\bar{3}d$ phase, provided that they are involved in two strong magnetic interactions: (i) a spin-orbit interaction, forcing the moments  to align along the easy magnetization axes coinciding with strings; (ii) an antiferromagnetic exchange between neighboring atoms on the  strings.  It is obvious that in this analogy the  conjugation plays the role of the time reversal operation, which in turn is independent of spatial symmetry. Therefore, it does not change the space group of the structure.

Note that the analogy is not complete, because, in contrast to  a magnetic moment, the  atomic position shift changes sign upon  inversion and  remains the same upon time  reversal. As for the strings, it is easy to see that they change their color upon rotations of $90^\circ$ and keep it unchanged for all other rotations of  point group $m\bar{3}m$, as well as for  inversion. For example, the $xyz$-component of a magnetic octupole moment behaves in a similar way. Using the magnetic analogy  we can extend the space group of a $bc8$-like phase by adding symmetry elements  conjugating its structure.  Then, the symmetry of the phase will be described by a Shubnikov magnetic group \cite{Litvin2013,Bilbao,Perez-Mato2015}.  Thus, the magnetic groups of the phases $bc8$, $bt8$, and $r8$  are $Ia\bar{3}d^\prime$, $I4_1/ac^\prime d^\prime$, and $R\bar{3}c^\prime$, correspondingly (Table~\ref{tab:phases}). In the supplemental material all $bc8$-like crystals are classified both by space and magnetic groups \cite{description}.

\section{Microscopic structure  and physical properties}
\label{sec:microscopic}

\begin{figure}[h]
\begin{center}
\includegraphics[width=7cm]{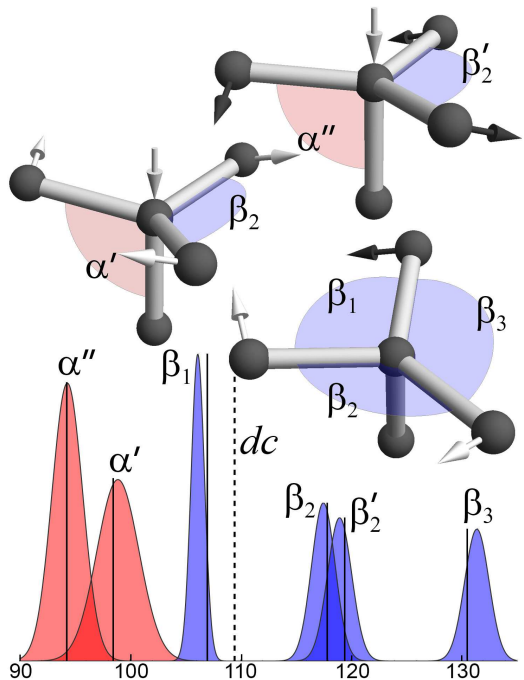}
\caption{\label{fig:angles} (Color online) The angles between interatomic bonds in the $bc8$-like structures depending on  switching of adjacent strings. The directions of switching are shown by the white/black arrows. The angles $\alpha$  are between bonds $A$ and $B$, and the angles $\beta$  are between $B$-bonds. In the plot, the Gaussians approximate statistical data on the real $bc8$-like crystals from Ref.~\cite{Rapp2015}. The vertical lines correspond to the ideal structure with the length of $A$-bonds equal to $\sqrt{2}/4\approx 0.35$  lattice parameters  of $bc8$. The dashed line  indicates the angle between  bonds in  diamond  structure, $\arccos(-1/3) \approx 109.5^\circ$.}
\end{center}
\end{figure}

Let us  now consider some structural features of the $bc8$-like phases and their correlation with physical properties such as energy and atomic density. As  mentioned  above, each atom has one $A$-bond  along the string passing through it, and three  $B$-bonds with atoms on other strings.  Therefore, two kinds of angles between the bonds can be distinguished, the $\alpha$ angle between bonds $A$ and $B$, and the $\beta$  angle between two $B$-bonds. In the real structures, all the bonds tend to be of the same size. It is achieved when the $A$- and $B$-bonds have lengths of about $\sqrt{2}/4\approx 0.35$ lattice parameters of $bc8$, and the $A^\prime$-distances are about $(2\sqrt{3}-\sqrt{2})/4\approx 0.51$ parameters of $bc8$. In such ideal structure, the values of $\alpha$ and $\beta$ are determined by  the colors  of the neighboring strings. Thus, depending on switching, the angles $\alpha$ are subdivided into $\alpha^\prime \approx 98.5^\circ$ and $\alpha^{\prime\prime} \approx 94.3^\circ$, and the angles $\beta$ into $\beta_1 \approx 107.0^\circ$, $\beta_2 \approx 117.9^\circ$, $\beta_2^\prime \approx 119.5^\circ$, and $\beta_3 \approx 130.6^\circ$ (Fig.~\ref{fig:angles}). The statistical variation of the angles calculated from the structural data from Ref.~\cite{Rapp2015} is shown  using Gaussian distributions. It is seen that the angles $\alpha$, $\beta_1$, $\beta_2$, and $\beta_3$ are well distinguished. Apparently, the angles $\alpha$, which are close to $90^\circ$, have  an excess energy, and the  structure should undergo an additional distortion in order to increase them.

\begin{figure}[h]
\begin{center}
\includegraphics[width=7cm]{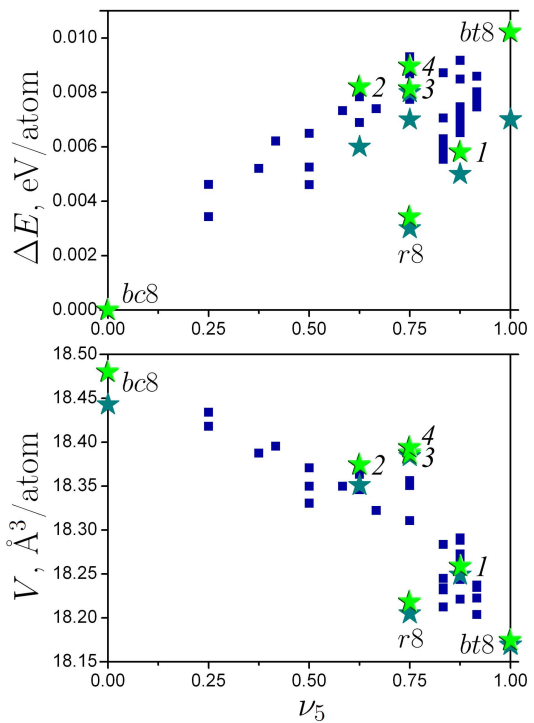}
\caption{\label{fig:E_V} (Color online) The energy and  volume per atom for $bc8$-like structures depending on the frequency  of five-membered rings calculated {\it ab initio}; the stars indicate experimentally observed structures, green for our results and dark cyan for the data from Ref.~\cite{Rapp2015}: {\sl 1} --- $m32$ ($m32$-8), {\sl 2} --- $m32^\star$ ($sm16$-2), {\sl 3} --- $t32$ ($t32$-1), {\sl 4} --- $t32^\star$ ($t32$-3); the squares  correspond to some new structures described in the present paper \cite{description}. The energy is measured from that of $bc8$.}
\end{center}
\end{figure}

An important characteristic of  tetrahedral structures is the statistics of atomic rings. All the considered $bc8$-like phases have  a girth (i.e.  the length of a shortest ring) equal to five, except for  $bc8$ itself, which is made up exclusively of six-membered rings. For the first approximation, we can investigate the dependence of  physical properties  on the  amount of five-membered rings per atom, which varies from $\nu_5=0$ for $bc8$ to $\nu_5=1$ for $bt8$. Fig.~\ref{fig:E_V} shows dependencies of the energy and volume per atom on the value of $\nu_5$  for several $bc8$-like phases, calculated during ab initio simulations of the structural relaxation of the phases, performed with {\sc Quantum ESPRESSO} package \cite{QE,QE-2009} (see details of the DFT modeling in the supplemental material \cite{description}). It  is seen that  when the frequency of five-membered rings grows,  both the energy and atomic density  increase.  The positive correlation between $\nu_5$ and energy per atom seems to be explained by the relation between five-membered rings and the ``bad'' angles $\alpha^{\prime\prime}$. Indeed, from geometrical considerations we can express the frequencies of different interbond angles through $\nu_5$: $\nu_{\alpha^\prime} = 3-2\nu_5$, $\nu_{\alpha^{\prime\prime}} = 2\nu_5$, $\nu_{\beta_1} = \nu_{\beta_2^\prime} = \nu_{\beta_3} = \nu_5$, $\nu_{\beta_2} = 3-3\nu_5$.

All $bc8$-like phases are close to each other but differ from other silicon phases. For example, the belonging of a given structure to the family can easily be determined by comparing its periods to those of the $bcc$ lattice, and by the number of atoms in the unit cell multiple of eight. Furthermore, all these phases have similar energy and atomic density. It was recently found that the Raman spectra of some unidentified silicon phases obtained in the experiment with the laser-induced microexplosions resemble the spectra of the phases $bc8$, $r8$, and $bt8$, although the correspondence is not exact \cite{Smillie2020}. Here, the existence of an infinite series of intermediate phases may be one of the possible explanations for the discrepancy.

\section{Discussion and conclusions}
\label{sec:conclusion}

In summary, we described in detail the physical mechanism behind the
complicated silicon structures observed by Rapp {\it et al.} after
ultrashort laser microexplosions \cite{Rapp2015}. Those silicon phases
(except $st12$) can be obtained from the phase $bc8$ by switching the
$\langle 111\rangle$ atomic strings in different regular (periodic) ways.
The stochastic switching of the strings gives a disordered phase with
$Ia\bar{3}d$ cubic symmetry, and the space groups of the $bc8$-like phases
are subgroups of $Ia\bar{3}d$. All the possible phase transitions between
different $bc8$-like phases should be of the first order (not continuous)
because atoms of the switched strings jump over finite distances.

We have found possible rich polymorphism of $bc8$-like phases, and to
distinguish between them it would be very important to study carefully any
small differences in diffraction patterns and in other physical properties
(e.g. in the Raman spectra). In addition, the considered family of
bc8-like phases provides a good opportunity for studying the ability of
existing empirical potentials to capture the structure and energetics of
these phases and other complicated silicon materials (thanks to the
Referee who attracted our attention to this option).

It would also be very interesting to use the technique of Rapp {\it et
al.} \cite{Rapp2015} for the $bc8$ crystals instead of diamond silicon
because in this case new $bc8$-like phases will grow on the parent $bc8$
matrix. Reasonably large (several mm size) phase-pure $bc8$ polycrystals
have been grown recently by different methods
\cite{Kurakevych2016,Li2019}.

Finally, we can propose a possible explanation for the observed
polymorphism of $bc8$-like phases based on the ideas described above. It
was shown in Ref.~\cite{Dmitrienko1999b}, that the reduced intensity
functions (the structure-dependent parts of the X-ray scattering pattern),
are very similar for amorphous silicon and for disordered polycrystalline
$Ia\bar{3}d$ phase. Thus, we can suppose that the
high-temperature/high-pressure amorphous phase first transforms into
disordered $Ia\bar{3}d$ phase and then, depending on local temperature,
pressure, and shear, into different $bc8$-like phases for which
$Ia\bar{3}d$ is the parent phase. Therefore, quite probably, the
$bc8$-like phases should also appear after the laser-induced
microexplosions on the interface between SiO$_2$ and amorphous silicon.

\section*{Acknowledgements}

We are grateful to S.~A.~Pikin, S.~M.~Stishov, V.~V.~Brazhkin,
M.~V.~Gorkunov, M.~Kl\'{e}man, V.~S.~Kraposhin, A.~L.~Talis, and
C.~J.~Pickard for useful discussions and communications. This work was
supported by the Ministry of Science and Higher Education of the Russian
Federation within the State assignment FSRC ``Crystallography and
Photonics'' RAS in part of symmetry analysis, and by the grant of
Prezidium of Russian Academy of Sciences in part of computer simulations.

\newpage

\onecolumngrid

\begin{table}
\caption{\label{tab:phases} The known $bc8$-like  phases of silicon}
\begin{tabular}{@{}l|l|l|l}
\hline
\hline
phase & Fedorov group & Shubnikov group & unit cell ideal periods \\
\hline
$bc8$ & $Ia\bar{3}$ (206) & $Ia\bar{3}d^\prime$ (230.148) & \multirow{2}{*}{$\left. \vphantom{\begin{array}{l} \\ \end{array}} \right\}$ $(100)$, $(010)$, $(001)$} \\
$bt8$ & $I4_1/a$ (88) & $I4_1/ac^\prime d^\prime$ (142.567) & \\
$r8$ & $R\bar{3}$ (148) & $R\bar{3}c^\prime$ (167.107) & $\left(\bar{\frac12}\frac12\frac12\right)$, $\left(\frac12\frac12\bar{\frac12}\right)$, $\left(\frac12\bar{\frac12}\frac12\right)$ \\
\hline
$m32$ ($m32$-8) & $P2_1/c$ (14) & $P2_1/c$ (14.75) & $\left(\frac12\frac12\bar{\frac12}\right)$, $(011)$, $\left(\frac32\bar{\frac12}\frac12\right)$ \\
$m32^\star$ ($sm16$-2) & $C2$ (5) & $C22^\prime 2^\prime$ (21.41) & $(100)$, $(011)$, $(01\bar{1})$ \\
$t32$ ($t32$-1) & $P\bar{4}2_1c$ (114) & $P_C\bar{4}2_1c$ (114.281) & \multirow{2}{*}{$\left. \vphantom{\begin{array}{l} \\ \end{array}} \right\}$ $(110)$, $(1\bar{1}0)$, $(001)$} \\
$t32^\star$ ($t32$-3) & $P4_32_12$ (96) & $P_C4_32_12$ (96.149) & \\
\hline
\hline
\end{tabular}
\end{table}

\begin{table}[h]
\caption{\label{tab:cells} The supercells multiple to the primitive cell of $bc8$ and the number of  possible $bc8$-like structures}
\begin{tabular}{@{}c|c|c|c|c}
\hline
\hline
cell & basis vectors & multiplicity & number of structures & the known \\
 &  &  &  &  \\
name & (periods of $bcc$ lattice) &  & without/with enantiomorphs & structures \\
 &  &  &  &  \\
\hline
8-1 & $(\frac12,\frac12,\bar{\frac12})$, $(\bar{\frac12},\frac12,\frac12)$, $(\frac12,\bar{\frac12},\frac12)$ & 1 & 3/3 & $bc8$, $r8$, $bt8$ \\
\hline
16-1 & $(\frac12,\frac12,\frac12)$, $(\frac12,\bar{\frac12},\frac12)$, $(1,0,\bar{1})$ & 2 & 6/8 & $m32^\star$ \\
16-2 & $(1,0,0)$, $(0,1,0)$, $(0,0,1)$ & 2 & --- & \\
\hline
24-1 & $(\frac12,\frac12,\frac12)$, $(\frac12,\bar{\frac12},\frac12)$, $(\frac32,\bar{\frac12},\bar{\frac32})$ & 3 & 18/23 & \\
24-2 & $(\frac12,\frac12,\frac12)$, $(0,\bar{1},1)$, $(1,0,\bar{1})$ & 3 & 4/4 & \\
24-3 & $(1,0,0)$, $(0,1,0)$, $(\frac12,\frac12,\frac32)$ & 3 & --- & \\
\hline
32-1 & $(1,0,0)$, $(0,1,1)$, $(0,\bar{1},1)$ & 4 & 6/8 & $t32$, $t32^\star$ \\
32-2 & $(\frac12,\frac12,\frac12)$, $(\frac12,\bar{\frac12},\frac12)$, $(2,0,\bar{2})$ & 4 & 46/68 & \\
32-3 & $(\frac12,\frac12,\frac12)$, $(1,0,\bar{1})$, $(\bar{\frac12},\frac32,\bar{\frac12})$ & 4 & 22/26 & $m32$ \\
32-4 & $(\frac32,\bar{\frac12},\bar{\frac12})$, $(\bar{\frac12},\frac32,\bar{\frac12})$, $(\bar{\frac12},\bar{\frac12},\frac32)$ & 4 & 26/40 & \\
32-5 & $(1,0,0)$, $(0,1,0)$, $(0,0,2)$ & 4 & --- & \\
32-6 & $(1,0,1)$, $(0,1,1)$, $(0,\bar{1},1)$ & 4 & --- & \\
32-7 & $(1,0,1)$, $(1,1,0)$, $(0,1,1)$ & 4 & --- & \\
32-8 & $(1,0,\bar{1})$, $(0,1,0)$, $(\frac12,\frac12,\frac32)$ & 4 & --- & \\
\hline
\multicolumn{3} {r|} {} & Total: 131/180 & \\
\hline \hline
\end{tabular}
\end{table}

\newpage
\twocolumngrid

\section*{SUPPLEMENTAL MATERIAL}

\section*{Animated GIFs}

Animation {\sl bondon.gif} demonstrates a probable mechanism of the string
switching from white to black and back. The process involves moving a
dangling interatomic bond along a 3-fold axis, which can be associated
with a qusiparticle called ``bondon''. Note that the sign of switching
does not correlate with the direction of bondon movement.

\section*{Crystallographic information files}

For generation and enumeration of possible $bc8$-like phases we use an
approach based on the "ideal" atomic structure of the $bc8$ phase as a
crystalline approximant of icosahedral quasicrystals where the switches of
the string correspond to phasonic jumps of atoms
\cite{Dmitrienko1999,Dmitrienko2000,Dmitrienko2001}. In this case, the $A$ and $B$
interatomic bonds are parallel, correspondingly, to three- and fivefold
axes of an icosahedron, with the ratio of bonds $r_B/r_A = \sqrt{(\tau +
2)/3} \approx 1.098$ ($\tau = (1+\sqrt{5})/2$ is the golden mean). When
the string is switched, the $A$ bonds remains of the same length and the
third type of bonds ($C$-bonds) appears, connecting atoms at switched and
non-switched strings.

The folder {\sl cifs\_.zip} contains Crystallographic Information Files
(CIFs) with atomic coordinates in such ideal representation. The cell
parameters are calculated based on the ideal $bc8$ crystal period $a =
6.658$\AA. Note that the crystal lattices of the idealized structures are
characterized by the unit cell parameters (angles, ratio of periods)
characteristic of the cubic lattice of the $bc8$ phase. A real structure
may have slightly distorted cell parameters if phase symmetry allows it.
The files from the folder {\sl cifs\_.zip} describe the structures as
having the least symmetrical group $P1$. All positions are listed for the
primitive cell. Some of the structures have been relaxed using {\sl ab
initio} simulations. In these cases, the cif-file, tagged by an extra
underscore symbol, also contains the refined crystallographic data. The
{\sl cifs\_.zip} file can be sent by request.

\section*{Details of ab initio simulation}

For the {\sl ab initio} DFT
relaxation of the initial ``ideal'' $bc8$-like structures we used the {\sc
Quantum ESPRESSO} code \cite{QE,QE-2009}. We selected the generalized
gradient approximation (GGA) with the Perdew--Burke--Ernzerhof (PBE)
exchange-correlation functional which seems to provide better agreement
between experimental and theoretical lattice parameters then the local
density approximation (LDA). We used the ultra-soft pseudopotential
{\sl Si.pbe-n-rrkjus\_psl.1.0.0.UPF} \cite{QE}, the plane-wave energy cutoff of
40 Ry, and the structural relaxations were supposed to be converged when
all of the interatomic forces were less than 10$^{-3}$ Ry/a.u. The
$bc8$-like structures relaxed this way are presented in the cif-files tagged by an extra
underscore symbol in the file names. This includes all $bc8$-like structures with the lattices 16-1, 24-1, 24-2, 32-1, the monoclinic phases $m32$-6,7,8 with the lattice 32-3, and the rhombohedral phases $r32$-3,4,5,6 with the lattice 32-4.

\section*{Lattices and structures}

Below 131 $bc8$-like phases are listed and classified (180 with chiral enantiomorphs). The list exhausts all similar structures with 8, 16, 24, and 32 atoms in primitive cells, including previously known ones ($bc8$, $bt8$, $r8$, $m32$-8 ($m32$ \cite{Rapp2015}), $sm16$-2 ($m32^\star$ \cite{Rapp2015}), $t32$-1 ($t32$ \cite{Rapp2015}), $t32$-3 ($t32^\star$ \cite{Rapp2015})), and 124 structures proposed for the first time.

The description of structures corresponds to the following scheme.

{\sl Lattice type.} The phases described are divided into eight different lattices: one each with 8 and 16 atoms in primitive cell, two with 24 atoms, and four with 32 atoms. Every lattice is defined by periods, which coincides with some periods of the $bcc$ lattice of the $bc8$ crystal. The type of Bravais lattice is indicated.

{\sl Symmetry.} The structures of $bc8$-like phases are classified by their symmetry. Crystals with the same symmetry are combined together. Listed are the space groups, the Pearson symbols, and the black-white magnetic groups (in red) of the structures.

{\sl Structures.} The name of each phase is constructed according to the
{\sl csN-n} scheme, where the optional symbol {\sl c} means centering
($b$: body-centered, $s$: base-centered), {\sl s} indicates crystal system
($a$: triclinic, $m$: monoclinic, $t$: tetragonal, $r$: rhombohedral, $h$:
hexagonal, $c$: cubic), {\sl N} is the number of atoms in the primitive
cell, {\sl n} is the sequence number. For a chiral structure, its
enantiomorph is indicated in parentheses.

\section*{LATTICE 8}

\noindent {\sc periods:} $a (\frac12,\frac12,\bar{\frac12})$, $a (\bar{\frac12},\frac12,\frac12)$, $a (\frac12,\bar{\frac12},\frac12)$

\vspace{0.2 cm}

\noindent {\sc bravais lattice:} body-centered cubic

\vspace{0.2 cm}

\noindent {\sc structures:}

\vspace{0.2 cm}

\noindent $Ia\bar{3}$ (206); $cI16$; \vd{$Ia\bar{3}d^\prime$ (230.148)} : $bc8$

\vspace{0.2 cm}

\noindent $I4_1/a$ (88); $tI16$; \vd{$I4_1/ac^\prime d^\prime$ (142.567)} : $bt8$

\vspace{0.2 cm}

\noindent $R\bar{3}$ (148); $hR24$; \vd{$R\bar{3}c^\prime$ (167.107)} : $r8$

\section*{LATTICE 16-1}

\noindent {\sc periods:} $a (\frac12,\frac12,\frac12)$, $a (\frac12,\bar{\frac12},\frac12)$, $a (1,0,\bar{1})$

\vspace{0.2 cm}

\noindent {\sc bravais lattice:} base-centered orthorhombic

\vspace{0.2 cm}

\noindent {\sc structures:}

\vspace{0.2 cm}

\noindent $C2$ (5); $mC32$; \vd{$C22^\prime2^\prime$ (21.41)} : $sm16$-1(2)

\vspace{0.2 cm}

\noindent $P2_1$ (4); $mP16$; \vd{$C2^\prime2^\prime2_1$ (20.33)} : $m16$-1(2)

\vspace{0.2 cm}

\noindent $P\bar{1}$ (2); $aP16$;

\hangindent=1cm \vd{$P2^\prime/c^\prime$ (13.69)} : $a16$-1, $a16$-2

\hangindent=1cm \vd{$P\bar{1}$ (2.4)} : $a16$-3, $a16$-4

\section*{LATTICE 24-1}

\noindent {\sc periods:} $a (\frac12,\frac12,\frac12)$, $a (\frac12,\bar{\frac12},\frac12)$, $a (\frac32,\bar{\frac12},\bar{\frac32})$

\vspace{0.2 cm}

\noindent {\sc bravais lattice:} face-centered orthorhombic

\vspace{0.2 cm}

\noindent {\sc structures:}

\vspace{0.2 cm}

\noindent $C2/c$ (15); $mC48$; \vd{$Fd^\prime d^\prime d$ (70.530)} : $sm24$-1, $sm24$-2

\vspace{0.2 cm}

\noindent $C2$ (5); $mC48$; \vd{$F2^\prime 2^\prime 2$ (22.47)} : $sm24$-3(5), $sm24$-4(6)

\vspace{0.2 cm}

\noindent $P\bar{1}$ (2); $aP24$;

\hangindent=1cm \vd{$C2^\prime/c^\prime$ (15.89)} : $a24$-1, $a24$-2, $a24$-3, $a24$-4, $a24$-5

\hangindent=1cm \vd{$C2^\prime/c^\prime$ (15.89)} : $a24$-6, $a24$-7

\hangindent=1cm \vd{$P\bar{1}$ (2.4)} : $a24$-8, $a24$-9, $a24$-10, $a24$-11

\vspace{0.2 cm}

\noindent\hangindent=1cm $P1$ (1); $aP24$; \vd{$C2^\prime$ (5.15)} : $a24$-12(13), $a24$-14(15), $a24$-16(17)

\section*{LATTICE 24-2}

\noindent {\sc periods:} $a (\frac12,\frac12,\frac12)$, $a (0,\bar{1},1)$, $a (1,0,\bar{1})$

\vspace{0.2 cm}

\noindent {\sc bravais lattice:} hexagonal

\vspace{0.2 cm}

\noindent {\sc structures:}

\vspace{0.2 cm}

\noindent $P\bar{3}$ (147); $hP24$; \vd{$P\bar{3}c^\prime1$ (165.95)} : $h24$-1, $h24$-2

\vspace{0.2 cm}

\noindent $P\bar{1}$ (2); $aP24$ : \vd{$C2^\prime/c^\prime$ (15.89)} : $a24$-18, $a24$-19

\section*{LATTICE 32-1}

\noindent {\sc periods:} $a (1,0,0)$, $a (0,1,1)$, $a (0,\bar{1},1)$

\vspace{0.2 cm}

\noindent {\sc bravais lattice:} tetragonal

\vspace{0.2 cm}

\noindent {\sc structures:}

\vspace{0.2 cm}

\noindent $P\bar{4}2_1c$ (114); $tP32$; \vd{$P_C\bar{4}2_1c$ (114.281)} : $t32$-1

\vspace{0.2 cm}

\noindent $P4_12_12$ (92); $tP32$; \vd{$P_C4_12_12$ (92.117)} : $t32$-2(3)

\vspace{0.2 cm}

\noindent $P4_32_12$ (96); $tP32$; \vd{$P_C4_32_12$ (96.149)} : $t32$-3(2)

\vspace{0.2 cm}

\noindent $C2/c$ (15); $mC64$; \vd{$C2/c$ (15.85)} : $sm32$-1, $sm32$-2

\vspace{0.2 cm}

\noindent $P\bar{1}$ (2); $aP32$; \vd{$C2^\prime/c^\prime$ (15.89)} : $a32$-1

\vspace{0.2 cm}

\noindent $P1$ (1); $aP32$; \vd{$P2^\prime$ (3.3)} : $a32$-2(3)

\section*{LATTICE 32-2}

\noindent {\sc periods:} $a (\frac12,\frac12,\frac12)$, $a (\frac12,\bar{\frac12},\frac12)$, $a (2,0,\bar{2})$

\vspace{0.2 cm}

\noindent {\sc bravais lattice:} base-centered orthorhombic

\vspace{0.2 cm}

\noindent {\sc structures:}

\vspace{0.2 cm}

\noindent\hangindent=1cm $C2$ (5); $mC64$; \vd{$C22^\prime 2^\prime$ (21.41)} : $sm32$-3(9), $sm32$-4(10), $sm32$-5(11), $sm32$-6(12), $sm32$-7(13), $sm32$-8(14)

\vspace{0.2 cm}

\noindent $P2_1$ (4); $mP32$; \vd{$C2^\prime 2^\prime 2_1$ (20.33)} : $m32$-1(3), $m32$-2(4)

\vspace{0.2 cm}

\noindent $P\bar{1}$ (2); $aP32$;

\hangindent=1cm \vd{$P2^\prime/c^\prime$ (13.69)} : $a32$-4, $a32$-5, $a32$-6, $a32$-7, $a32$-8, $a32$-9, $a32$-10, $a32$-11, $a32$-12, $a32$-13, $a32$-14, $a32$-15

\hangindent=1cm \vd{$P\bar{1}$ (2.4)} : $a32$-16, $a32$-17, $a32$-18, $a32$-19, $a32$-20, $a32$-21, $a32$-22, $a32$-23, $a32$-24, $a32$-25, $a32$-26, $a32$-27

\vspace{0.2 cm}

\noindent $P1$ (1); $aP32$;

\hangindent=1cm \vd{$C2^\prime$ (5.15)} : $a32$-28(30), $a32$-29(31), $a32$-32(34), $a32$-33(35)

\hangindent=1cm \vd{$P2^\prime$ (3.3)} : $a32$-36(42), $a32$-37(43), $a32$-38(44), $a32$-39(45), $a32$-40(46), $a32$-41(47)

\hangindent=1cm \vd{$P1$ (1.1)} : $a32$-48(52), $a32$-49(53), $a32$-50(54), $a32$-51(55)

\section*{LATTICE 32-3}

\noindent {\sc periods:} $a (\frac12,\frac12,\frac12)$, $a (1,0,\bar{1})$, $a (\bar{\frac12},\frac32,\bar{\frac12})$

\vspace{0.2 cm}

\noindent {\sc bravais lattice:} monoclinic

\vspace{0.2 cm}

\noindent {\sc structures:}

\vspace{0.2 cm}

\noindent\hangindent=1cm $P2_1/c$ (14); $mP32$; \vd{$P2_1/c$ (14.75)} : $m32$-5, $m32$-6, $m32$-7, $m32$-8

\vspace{0.2 cm}

\noindent $P\bar{1}$ (2); $aP32$;

\hangindent=1cm \vd{$P2^\prime/c^\prime$ (13.69)} : $a32$-56, $a32$-57, $a32$-58, $a32$-59

\hangindent=1cm \vd{$P2_1^\prime/c^\prime$ (14.79)} : $a32$-60, $a32$-61

\hangindent=1cm \vd{$P\bar{1}$ (2.4)} : $a32$-62, $a32$-63, $a32$-64, $a32$-65, $a32$-66, $a32$-67, $a32$-68, $a32$-69

\vspace{0.2 cm}

\noindent $P1$ (1); $aP32$;

\hangindent=1cm \vd{$P2^\prime$ (3.3)} : $a32$-70(72), $a32$-71(73)

\hangindent=1cm \vd{$P1$ (1.1)} : $a32$-74(76), $a32$-75(77)

\section*{LATTICE 32-4}

\noindent {\sc periods:} $a (\frac32,\bar{\frac12},\bar{\frac12})$, $a (\bar{\frac12},\frac32,\bar{\frac12})$, $a (\bar{\frac12},\bar{\frac12},\frac32)$

\vspace{0.2 cm}

\noindent {\sc bravais lattice:} rhombohedral

\vspace{0.2 cm}

\noindent {\sc structures:}

\vspace{0.2 cm}

\noindent $R\bar{3}$ (148); $hR96$; \vd{$R\bar{3}c^\prime$ (167.107)} : $r32$-1, $r32$-2

\vspace{0.2 cm}

\noindent $R3$ (146); $hR96$; \vd{$R32^\prime$ (155.47)} : $r32$-3(5), $r32$-4(6)

\vspace{0.2 cm}

\noindent $P\bar{1}$ (2); $aP32$;

\hangindent=1cm \vd{$C2^\prime/c^\prime$ (15.89)} : $a32$-78, $a32$-79, $a32$-80, $a32$-81, $a32$-82, $a32$-83

\hangindent=1cm \vd{$P\bar{1}$ (2.4)} : $a32$-84, $a32$-85, $a32$-86, $a32$-87

\vspace{0.2 cm}

\noindent $P1$ (1); $aP32$

\hangindent=1cm \vd{$Cc^\prime$ (9.39)} : $a32$-88(90), $a32$-89(91)

\hangindent=1cm \vd{$C2^\prime$ (5.15)} : $a32$-92(98), $a32$-93(99), $a32$-94(100), $a32$-95(101), $a32$-96(102), $a32$-97(103)

\hangindent=1cm \vd{$P1$ (1.1)} : $a32$-104(108), $a32$-105(109), $a32$-106(110), $a32$-107(111)

\end{document}